\documentclass[review]{elsarticle}

\usepackage{amssymb}
\usepackage{amsmath}
\usepackage{color}
\usepackage{caption,booktabs,array}

\usepackage{lineno,hyperref}
\modulolinenumbers[5]
\usepackage{float}
\journal{Meat science}

\biboptions{authoryear}

\usepackage{changepage}
\usepackage{booktabs,makecell,multirow}
\usepackage{xfrac}
\usepackage[group-digits=false]{siunitx}
\usepackage{amsmath}

\usepackage{geometry}

\begin{document}
\begin{frontmatter}

\title{Towards real time assessment of intramuscular fat content in meat using optical fibre-based optical coherence tomography}

\author[a,b]{Abi Thampi}
\ead[url]{https://www.biophotonics-newzealand.com/}
\cortext[mycorrespondingauthor]{Corresponding author}
\ead{atha302@aucklanduni.ac.nz}

\author[b,c]{Sam Hitchman}
\author[a,b]{St\'{e}phane Coen}
\author[a,b]{Fr\'{e}d\'{e}rique Vanholsbeeck}

\address[a]{Department of Physics, The University of Auckland,1010, New Zealand}
\address[b]{The Dodd-Walls Centre for Photonic and Quantum Technologies, New Zealand}
\address[c]{Agresearch, Hamilton, New Zealand} 

\begin{abstract}
  We consider the use of optical coherence tomography (OCT) imaging to predict the quality of meat. We find that intramuscular fat (IMF) absorbs infrared light about nine times stronger than muscle, which enables us to estimate fat content in intact meat samples. The method is made very efficient by extracting relevant information from the three-dimensional high-resolution images generated by OCT using principal component analysis (PCA). The principal components are then used as regressors into a support vector regression (SVR) prediction model. The SVR model is found to predict IMF content stably and accurately, with an $R^2$~value of~$0.94$. Our study paves the way for automated, contact-less, non-destructive, real time classification of the quality of meat samples.     
\end{abstract}


\end{frontmatter}


\section{Introduction}

\noindent Intramuscular fat (IMF) is an important quality attribute of meat that determines the eating experience of customers. In general, it positively influences juiciness, tenderness, flavour and the overall acceptability of meat in different breeds~(\cite{hocquette2010intramuscular}). 

The meat industry faces challenges in marketing meat products according to their qualities due to the high variability and heterogeneity of samples~(\cite{hocquette2010intramuscular}). Standard quality assessment methods, chemical or mechanical, are often destructive and slow, have huge sampling errors, and require large expenditure on personnel and equipment. Moreover, they are often not adaptable to an ``in-line'' usage. In particular, internationally recognized methods for IMF analysis, such as Soxhlet solvent extraction, the Folch method, or gravimetric measurements of the extracted fat weight~(\cite{silva2015intramuscular}), all have to be performed in a dedicated laboratory, outside the meat processing plant, and require minced meat for analysis. Overall, the evaluation, prediction, and control of quality attributes of individual pieces of meat in processing plants remains elusive.

To address these challenges, a number of techniques have been investigated over the past few years for application to meat quality attribute measurements, including visible or near-infrared (Vis-NIR) spectroscopy~(\cite{prieto2009application}), hyper-spectral imaging~(\cite{feng2018hyperspectral,lohumi2016application}), and ultrasound imaging~(\cite{simal2003ultrasonic}). These techniques are non destructive, but still have severe drawbacks. For a start, they mostly only determine an average composition across an entire sample, as there is in general a trade off between spectral resolution and imaging speed. The low spatial resolution of Vis-NIR spectroscopy, e.g.\ is a disadvantage when analysing highly non-homogeneous meat samples such as whole steaks. Hyper-spectral cameras are also limited to surface examination. Ultrasound imaging can probe deep inside samples but requires direct contact. Finally, hyper-spectral cameras and Vis-NIR spectroscopy both require an external light source, making calibration and comparison across processing plants difficult in the long term.  

Clearly, there is a need for a fast, reproducible, and non-destructive method to predict meat quality in real time. In this work, we consider optical coherence tomography (OCT) and show that this contact-less, three-dimensional, high-resolution imaging modality holds great promises for this task. OCT can be understood as a light-based version of ultrasound imaging. It generates images by measuring the intensity of light back-scattered from the depth of a sample. Unlike ultrasound where the delay between the original sound and its echoes can be directly measured electronically, OCT uses white-light (low-coherence) interferometry to measure the time of flight from reflections (\cite{boudoux2016fundamentals}). OCT produces a tomographic map of a sample by moving the light beam in a raster pattern. The depth profile at each location is called an A-scan (Axial). A line of A-scans form a B-scan, which is a 2D slice through the sample. A series of B-scans, separated by a given distance, produces a volume or C-scan, which shows the structure of the measured sample volume. Typical OCT systems that provide a lateral resolution of 5--20~$\mu$m, and kHz to MHz A-scan acquisition rates, are nowadays commercially available for fast volume imaging, and commonly applied to semi-transparent tissues, such as the skin, eyes, and some internal organs (\cite{bouma2001handbook}).

Continuous high resolution imaging of meat samples with OCT can in principle be implemented in a processing plant setting and will provide a lot of detailed information. However, the amount of data generated, combined with the large spatial variability of the samples~(\cite{chen2014deep}), significantly complicates classification. To address this problem, we combine OCT with two machine learning techniques. As a first step, we use principal component analysis (PCA) to reduce the dimensions of the complex imaging data set. PCA projects a set of observations to a new space spanned by a few number of dimensions, or components, orthogonal and uncorrelated to each other in such a way that each successive component accounts for a decreasing amount of variability in the data. This procedure filters out noise and reveals hidden structure in the data~(\cite{shlens2014tutorial}). For each sample, the first few principal components are then used as input variables (regressors) in a regression model for classification. That model is determined by another machine learning technique, namely support-vector regression (SVR), which has emerged as an elegant tool for solving regression problems based on a training data set~(\cite{vapnik1999overview, dibike2001model, drucker1997support, basak2007support}). It is a form of supervised learning, also known as support-vector machine. Reducing the number of variables through PCA prior to SVR helps to make the overall classification very efficient. 

In this paper, we test this approach by analysing small sections of beef shortloin from Wagyu and Friesian bull animals to make and test the regression models, and we show that PCA-based feature extraction of OCT images is a viable alternative for fast prediction of meat quality, with the potential for real time implementation in the near future. 

\section{Materials and methods}
\subsection{Meat samples}

\noindent \emph{M. longissimus thoracis et lumborum} (striploin) comprising of Wagyu X Friesian (n = 52) and Friesian bull (n = 130) were collected from a local abattoir during three processing days. The Wagyu samples were obtained by excising a single striploin from animals across a range of marble scores to ensure a wide range of IMF values (Scored 2--6) at approximately 16~hours post-mortem. The striploins were packaged individually in unsealed vacuum bags and placed in cartons and held at $-1.5\ ^\circ$C until approximately 24~hours post-mortem. The Friesian bull samples were obtained from a hot-boning process at a local abattoir, with both the left and right striploins being excised approximately 30~minutes post-mortem and vacuum packed in accordance with normal plant operation. 

\begin{figure}[t]
  \centering
  \includegraphics[width=0.93\textwidth]{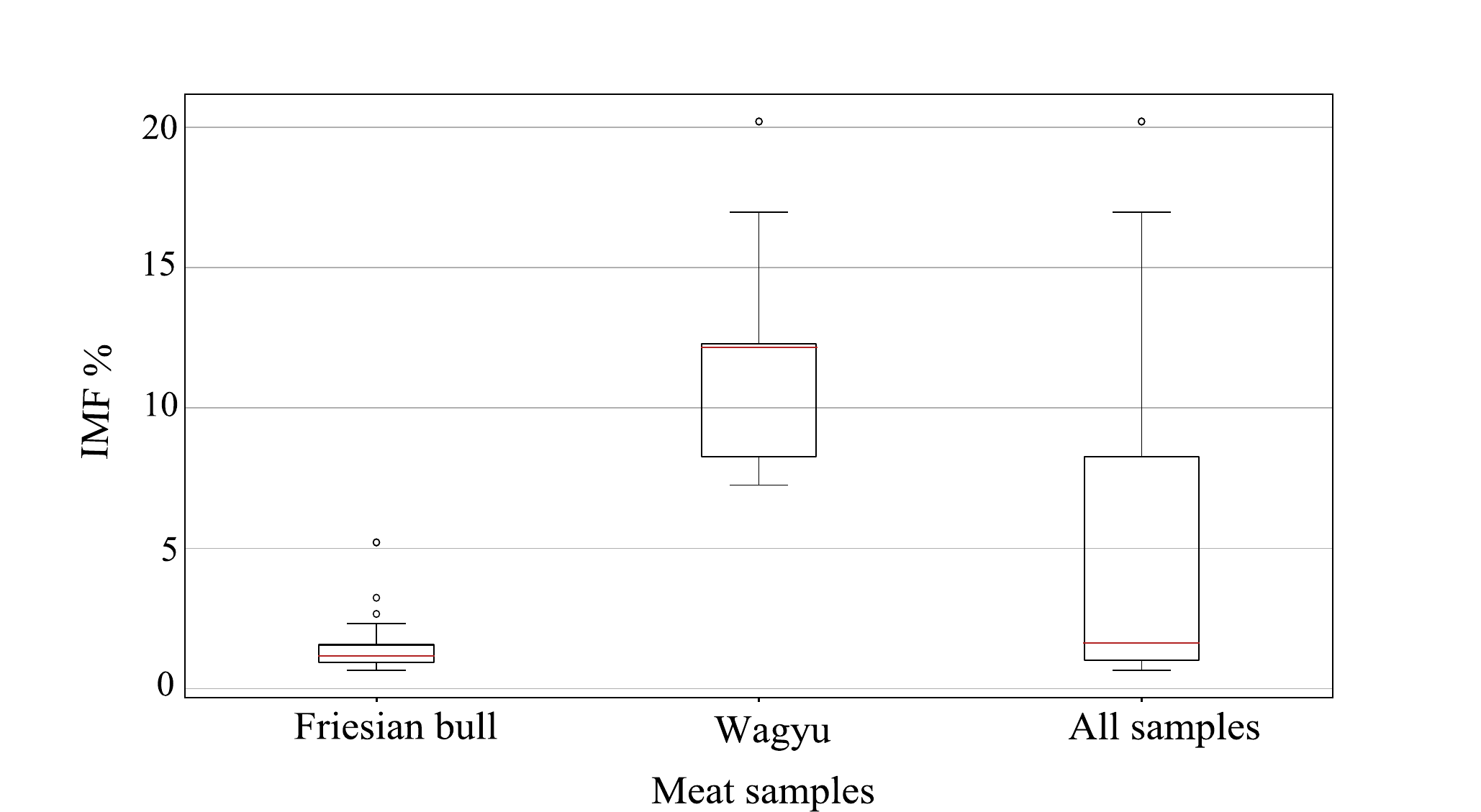}
  \caption{IMF\,\% distribution versus type of samples used in this study. The IMF\,\% is measured after OCT imaging using gas chromatography as per industry standards.}
  \label{fig:samples}
\end{figure}

At approximately 24 hours post-mortem, samples were cut into five steaks each. 70 of these samples --- Wagyu (n=24) and Friesian bull (n=46) --- were allocated to our OCT study. The remaining samples were used in other studies and are not considered here~(\cite{sam2020}). All sides (n=6) of each of the 70~samples were imaged on the same day at room temperature ($20\,^\circ$C). After OCT imaging, the steaks were freeze dried and the fatty acid composition was measured using gas chromatography as outlined in~\cite{craigie2017application} to provide comparison data. Figure~\ref{fig:samples} shows the distribution of IMF\,\% of the Friesian bull (n=46), Wagyu (n=24), and total (n=70) dataset.   

\subsection{OCT system}

\begin{figure}[t]
  \centering
  \includegraphics[width=0.9\linewidth]{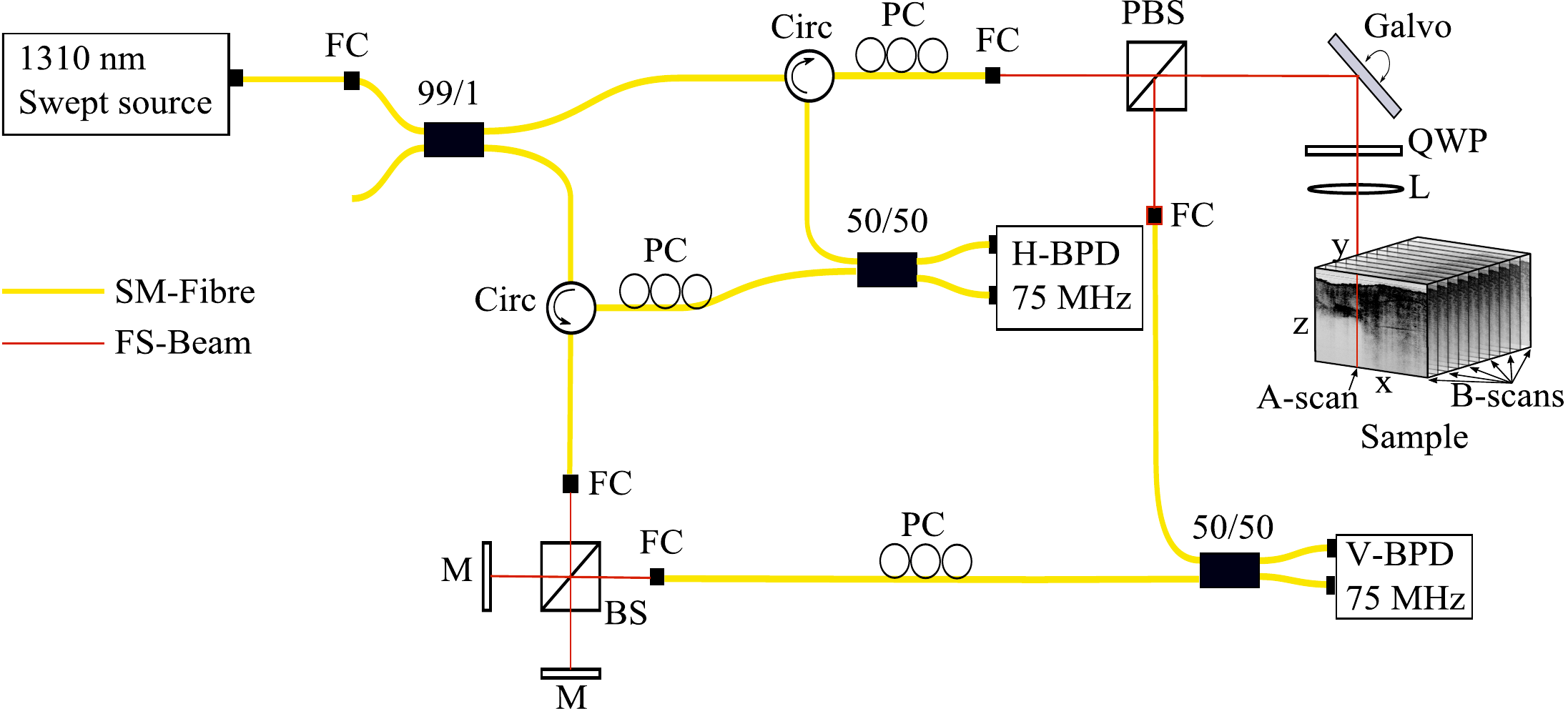}
  \caption{Diagram of the OCT setup. Single-mode (SM) optical fibers are highlighted in yellow, while free space (FS) parts are shown in red. 1310~nm: laser light source, 99/1 and 50/50: fiber couplers, Circ: circulator, FC: fiber collimator, L: lens, M: mirror, Galvo: galvanometer-controlled rotating mirror, BS: 50/50 beam splitter, PBS: polarization beam splitter, PC: polarization controller, QWP: quarter-wave plate, BPD: balanced photo detector.}
  \label{fig:setup_SM}
\end{figure}

\noindent Measurements are taken using a swept-source OCT system as shown in Fig.~\ref{fig:setup_SM}. This system is custom-built using a combination of single mode optical fibers and free space optics but the overall configuration otherwise follows standard OCT practice (\cite{fujimoto2000optical}). The light source is a commercially available wavelength-swept laser (Axsun Technologies Inc, Billerica, Massachusetts) with a central wavelength of~1310~nm, bandwidth of 100~nm, 50~kHz sweep rate, and an average output power of 33~mW. Light back-scattered by the sample is mixed with a reference beam and the resulting interference intensity patterns are detected by balanced photo detectors (BPD). This is done separately for the two orthogonal polarization directions (respectively V and H), which are split with a polarization beam-splitter (PBS). The system is thus polarization-sensitive and in principle able to detect sample birefringence, but this feature was not used in the present study. Rather, we only considered the total combined intensity of the two balanced photo detectors as our OCT signal. 

As the wavelength is swept, standard processing of the OCT signal leads to A-scans, which are effectively a measure of the total backscattered light reflectivity $R(z)$ at the beam position as a function of depth~$z$ inside the sample. B-scans and C-scans are then constructed by raster scanning the laser beam across the sample using two rotating mirrors, each controlled by a galvanometer (only one is shown in Fig.~\ref{fig:setup_SM}). There are 714 A-scans in a B-scan (along what we define as the $x$-axis), and 150 B-scans in a C-scan (respectively, $y$-axis). Each C-scan taken by our system represents a volumetric image of 1~cm $\times$ 1~cm ($x$--$y$ cross-section) $\times$ 2.5~mm (depth,~$z$). Several of such C-scans are measured in a grid-like pattern for each face of our 70 samples.



\subsection{Attenuation calculations}

\noindent The total backscattered light reflectivity $R(z)$ is related to 
the reflectance $\rho$ of a scatterer at depth~$z$ in the sample and to the attenuation coefficient $\mu_\mathrm{t}(z)$ according to the Beer-Lambert law (in the single scattering approximation)  (\cite{schmitt1993measurement}),
\begin{equation}
  R(z) = \rho \, \exp \left(-2 \int_0^z \mu_\mathrm{t}(z) dz \right)\,.
  \label{eq:attenuation}
\end{equation}
The Beer's law relates intensity of light passing through a section of sample to the thickness of the sample as a simple exponential decay of the light incident on the sample. The attenuation coefficient $\mu_\mathrm{t}$~$(\mathrm{mm}^{-1}$) describes this decay. The factor of~2 in the exponential term accounts for the double pass of the light, in and out of the sample. 


By taking the  natural logarithm of $R(z)$ in Equation~(\ref{eq:attenuation}) and differentiating the result, we obtain
\begin{equation}
    \frac{d \ln{R(z)}}{dz} = - 2\mu_\mathrm{t}(z)\,.
    \label{eq:logR}
\end{equation}
Equation~(\ref{eq:logR}) shows that the gradient of $\ln{R(z)}$ is proportional to the local value of the absorption coefficient~$\mu_\mathrm{t}$. This quantity can thus be obtained very simply from A-scans as a function of depth. As $\mu_\mathrm{t}(z)$ is dependent on the sample optical properties, it can be used to differentiate intramuscular fat from muscle.

\begin{figure}[t]
  \centering
  \includegraphics[scale=0.60]{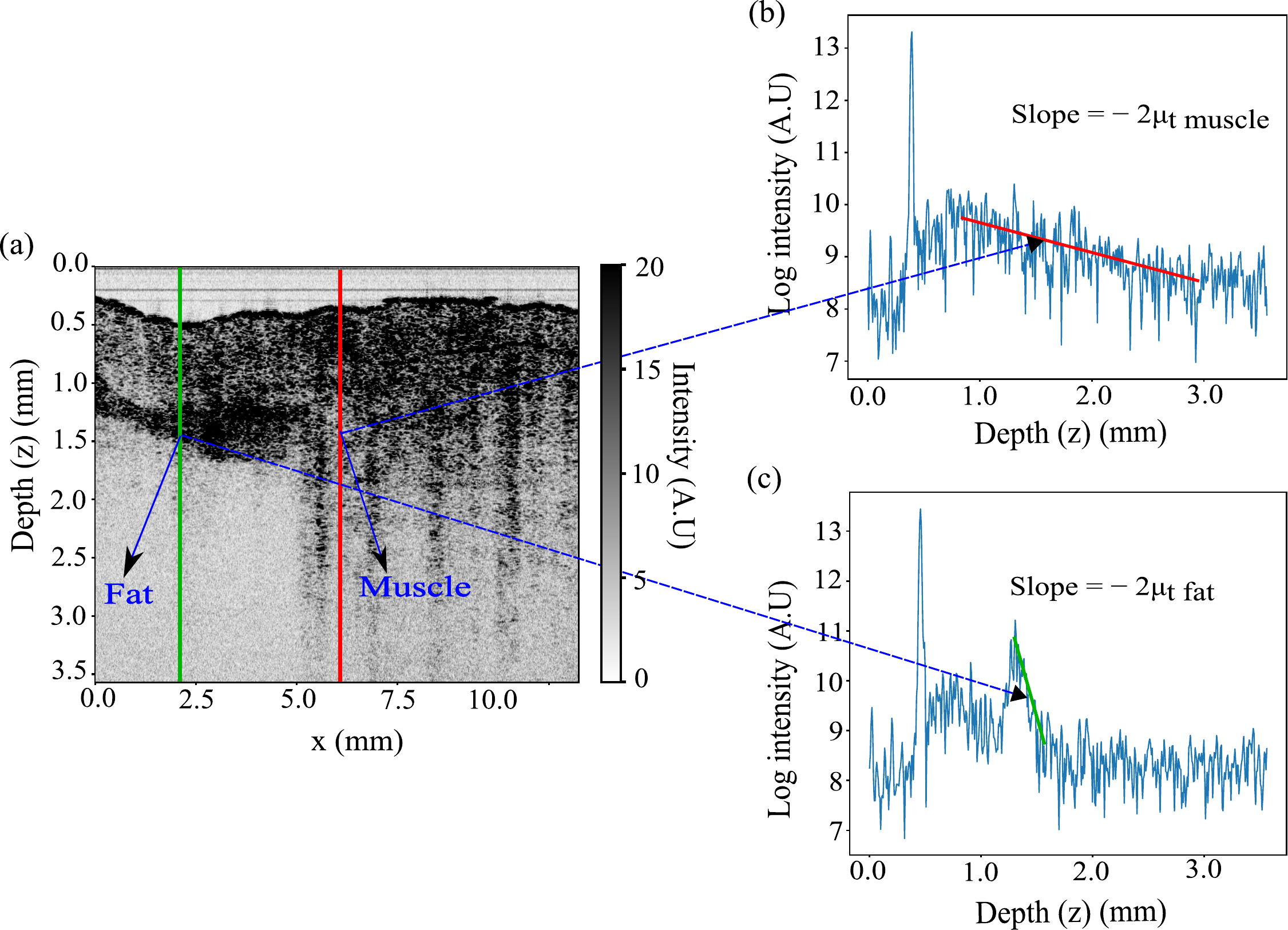}
  \caption{(a) B-scan of a meat sample containing muscle fibre and IMF. (b), (c) Individual A-scans corresponding to the red and green vertical lines in (a), plotted as natural logarithm of the OCT signal. The slope of the line of best fit of the selected features represent the attenuation coefficient $\mu_\mathrm{t}$ of muscle and fat, respectively.}
  \label{fig:atten_curve}
\end{figure}

We illustrate this aspect in Fig.~\ref{fig:atten_curve}. Panel~(a) shows one B-scan of a sample containing some IMF along with meat muscle. In this image, we have selected two A-scans, respectively in areas with only muscle (highlighted by the red line), and with a mixture of fat and muscle (green line). The natural logarithm of the corresponding OCT signals are plotted in Figs~\ref{fig:atten_curve}(b) and~(c) respectively. As shown, the gradient of the line of best fit of the features shown in these plots enable us to estimate the absorption coefficient~$\mu_\mathrm{t}$ of muscle and IMF. 

\subsection{Attenuation heat-map on OCT data}

\noindent To automate the analysis of the attenuation, an attenuation ``heat-map'' is built for each C-scan by downsampling the data. Specifically, we define a grid of small voxels across the C-scans, with voxels large enough to span across multiple A-scans. We first find the gradient of the line of best fit for each section of A-scan data in each voxels following the procedure above to extract the attenuation coefficient. Gradient values are then averaged out across voxels to give an \emph{effective} average attenuation coefficient, $\langle\mu_\mathrm{t}\rangle$, for each voxel. We use the word `effective' here because in contrast to a real absorption coefficient (which is always positive) $\langle\mu_\mathrm{t}\rangle$ can take both signs: positive for absorbing regions as for the sections highlighted in Figs~\ref{fig:atten_curve}(b)--(c), and negative at interfaces between muscle and IMF, due to light being reflected by refractive index discontinuities [see, e.g.\ Fig.~\ref{fig:atten_curve}(c)]. While the negative average gradients could be considered artefacts, they still carry some information about the sample which is beneficial to our machine learning approach. Therefore, we do not need to filter those values out, which simplifies signal processing.

\begin{figure}[t]
  \centering
  \includegraphics[width=1\linewidth]{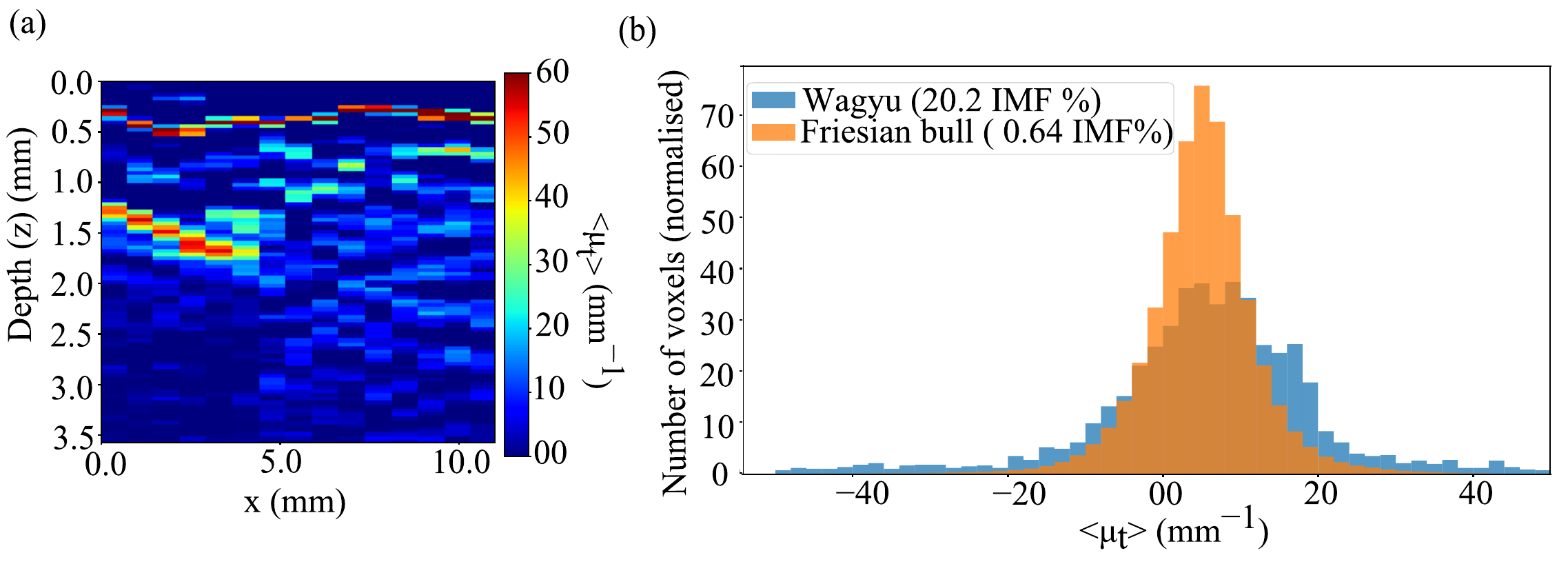}
  \caption{(a) Heat-map of effective averaged attenuation coefficient $\langle\mu_\mathrm{t}\rangle$, corresponding to the B-scan shown in Fig.~\ref{fig:atten_curve}(a). Voxels with high attenuation are considered to be 100\,\% IMF. The apparent high attenuation of the top surface is an artefact due to the large amount of light reflected at the air-meat interface. (b)~Histograms of effective attenuation of two representative samples, respectively with high (Wagyu, blue) and low (Friesian bull, orange) IMF\,\%.}
  \label{fig:histogram}
\end{figure}

As an example, we show in Fig.~\ref{fig:histogram}(a) the attenuation heat-map corresponding to the B-scan plotted in Fig.~\ref{fig:atten_curve}(a). Here, we only show positive values for better contrast. Red pixels are associated with high attenuation and therefore represent fat. We can observe that large fat deposits (such as on the left) absorb all the light, shadowing any structures underneath. To take that into account in our analysis, we automatically remove data with a signal-to-noise ratio (SNR) below a certain threshold, limiting the depth of the images. The sections above the surface of the samples, including the initial reflection artifact, are similarly removed. Note that the size of voxels was determined empirically: large enough to improve SNR and get enough information to predict IMF\,\%, but not exceeding the size of marbling, which is typically in the~mm range. 


\subsection{Histogram and distribution of IMF in the samples}

\noindent To increase processing speed and further reduce the complexity and dimensions of the data, we  convert the heat-maps into histograms. Two representative histograms are shown in Fig.~\ref{fig:histogram}(b) for, respectively, high fat (Wagyu, blue) and low fat (Friesian bull, orange) samples. As should be clear from the discussion above, the large positive values of~$\langle\mu_\mathrm{t}\rangle$ relate to voxels which mainly contain IMF, while the negative values represent the interfaces between IMF and meat muscle. A meat sample with a large IMF content is therefore expected to have more voxels towards both the high negative and positive sides of the histogram of effective attenuation coefficients, i.e.\ to be more spread-out. Fig.~\ref{fig:histogram}(b) confirms this interpretation.


Representing our data in terms of histograms offer a one-dimensional view  of the entire fat content across the volume of a sample. We argue that such histograms are more effective in determining IMF\,\% than counting, in the heat-maps, the individual voxels associated with IMF or meat muscle. This latter method lead to ambiguity for voxels containing a mixture of muscle and IMF and which present intermediate attenuation values. This occurs in particular at interfaces between muscle and IMF. Our histogram representation handles such cases graciously without discarding any voxel data. Overall, histograms offer a more faithful picture of the fat content of a sample. One-dimensional data-sets such as histograms are also more adapted to dimension reduction tools like PCA and spatial-spectral regression techniques like SVR, leading to better prediction models. Note that because IMF is distributed non-uniformly in each meat sample, the histograms we use in our analysis include data from all C-scans collected from a sample.

\subsection{Machine learning techniques}

\noindent Principal Component Analysis (PCA) is used to reduce the number of independent variables in the histograms (bins = 200) to build an effective and robust regression model~(\cite{balabin2011support}). The histograms of the samples are normalised to the area under the curve prior to PCA. The principal components (PCs) obtained from the PCA output are used as regressors for the SVR analysis while the IMF\,\% obtained using industry standards (Fig.~\ref{fig:samples}) are used as the dependent variables. 

The SVR model was developed using 55~samples for the training data set, and was used to predict the IMF\,\% of the remaining 15~samples (test set). The samples were divided randomly into training and test sets. Test sets had at least one random sample from each IMF\,\% range. The 15~samples in a test set consisted of 3~samples with 0--1~IMF\,\%, 4~samples with 1--2~IMF\,\%, 2~samples with 2--5~IMF\,\%, 3~samples with 5--10\,IMF\,\%, 2~samples with 10--15~IMF\,\%, and 1~sample with 16--20~IMF\,\%. The PCA model was built on the training set and the associated loadings were used to make the SVR model. PCs were then calculated for each sample in the test set, based on the PCA loadings obtained from training data. The test data was completely unknown to the algorithm making the PCA and regression model, and therefore can be reliably used to assess the accuracy of the model. Cross validation was done five times by randomly picking up different test and training sets from the overall data set. 

All the analysis performed in our study was done in standard Python~3.6, with the Scikit-learn library for the PCA and SVR algorithms. 

\section{Results}

\subsection{Heat-maps and attenuation histograms}

\noindent To build the attenuation heat maps, voxel size was chosen as large as possible to optimize computational speed without degrading the PCA grouping. This led to voxels  600~$\mu$m (along~$x$) $\times$ 285~$\mu$m (along~$y$) $\times$ 40~$\mu$m (depth,~$z$). From the heat maps we have also quantified the difference in attenuation between fat and muscle. Fat scatters light more than muscle resulting in stronger light attenuation, as shown by the steeper slope highlighted in Fig.~\ref{fig:atten_curve}(c) in comparison to that seen in Fig.~\ref{fig:atten_curve}(b). The average attenuation coefficients of muscle and fat were found as:
\begin{align*}
  \mu_\mathrm{t}(\mathrm{muscle}) &= (7.05 \pm 5.13)\ \mathrm{mm}^{-1}\\
  \mu_\mathrm{t}(\mathrm{fat}) &= (62.06 \pm 20.32)\ \mathrm{mm}^{-1}
\end{align*}
Clearly, the difference is significant and underlies our ability to characterise IMF\,\% with OCT.


\subsection{PCA results}

\begin{figure}[t]
  \centering
  \includegraphics[width=1.0\linewidth]{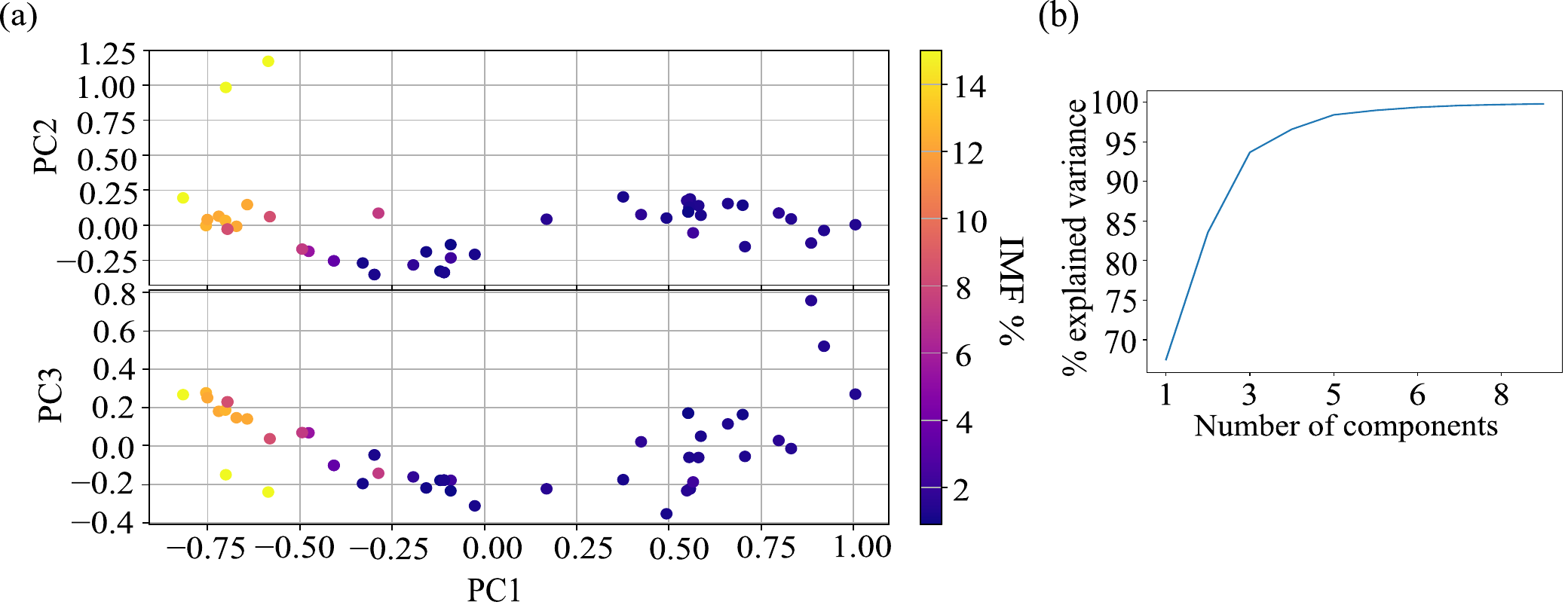}
  \caption{(a) PCA analysis of the attenuation histograms of different samples. The samples with high IMF\,\% (yellow tinted dots) clearly appear grouped together and separated from low IMF\,\% samples (blue tinted dots). The IMF\,\% associated with the colour scale was determined from subsequent gas chromatography analysis of the samples. (b)~The \%-explained variance curve shows how much information can be attributed to each PCs.}
  \label{fig:PCA_2d}
\end{figure}

\noindent Normalized histograms show clear differences between low and high IMF\,\% samples as was shown in Fig.~\ref{fig:histogram}(b). The PCA analysis of the histograms was able to successfully group the samples according to high IMF\,\% and low IMF\,\%. This is illustrated in Fig.~\ref{fig:PCA_2d}(a) in which the points associated with high fat samples (yellow tint) are grouped together and clearly separated from the low fat ones (blue tint). We found that we could limit the PCA analysis to only five principal components (PCs), as these were sufficient to account for $98.4$\,\% of the variance in the histograms of different samples as shown in Fig.~\ref{fig:PCA_2d}(b).

\subsection{Modelling IMF\,\% using SVR}

\noindent For the SVR modelling, the first five PCs of the PCA model are used as the regressors, or input variables, while the IMF\,\% obtained from gas chromatography is the dependent variable. The model was tested with a number of PCs from~1 to~6, but five PCs was found to lead to the best outcome.

\begin{figure}[ht]
  \centering
  \includegraphics[width=0.8\linewidth]{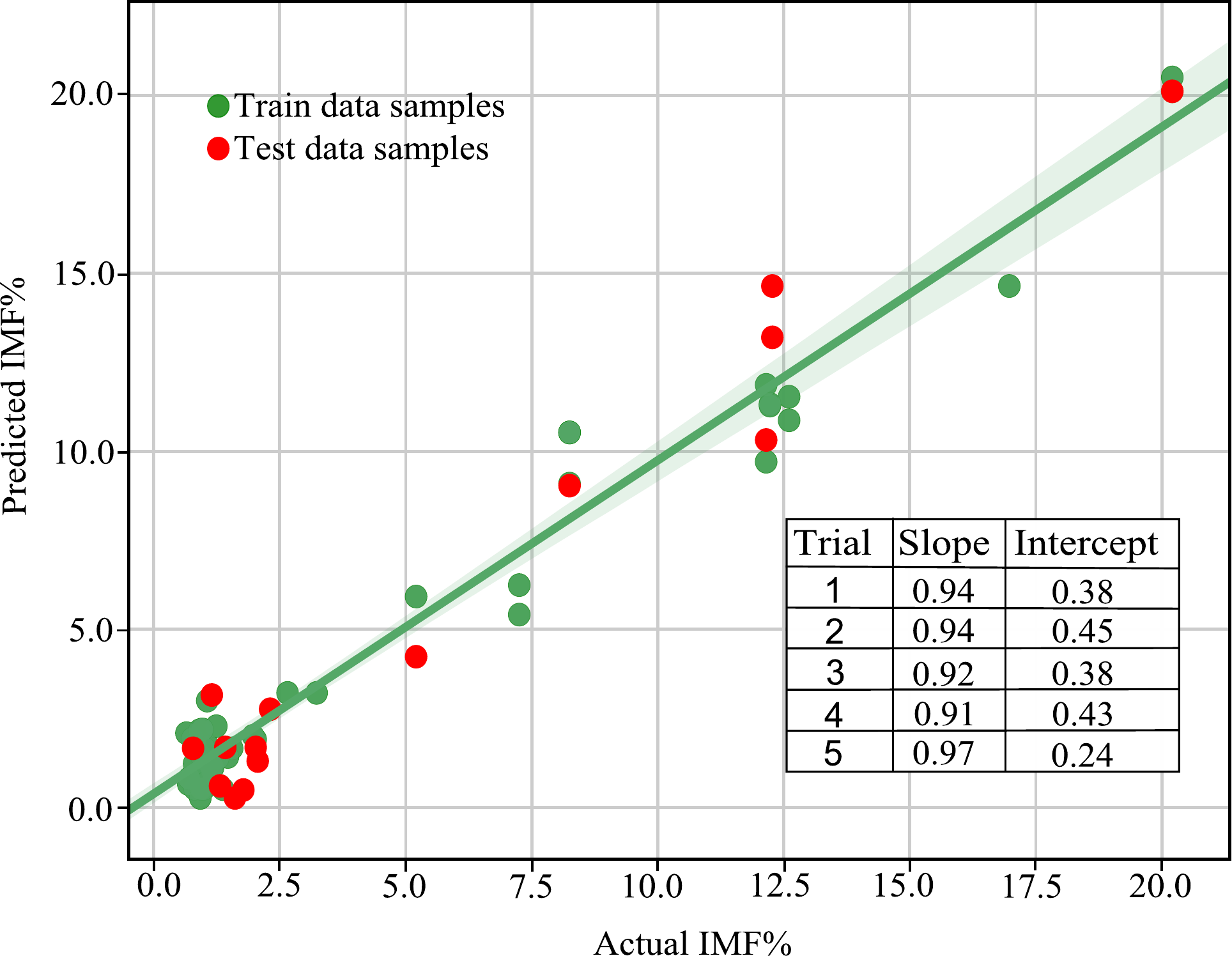}
  \caption{IMF\,\% predicted by our SVR model versus actual values for the training (green points) and test (red points) samples of Trial~2. The green line is a fit across the training points, along with the 95\,\% confidence interval (translucent green). The table shows the slope and intercept (which ideally should be 1 and~0, respectively) of the regression model fit for each of the five trials.}
\label{fig:SVR}
\end{figure}

As shown in Table~\ref{tab:SVR}, the SVR prediction model shows good stability  and prediction accuracy over several trials (using different training sets) with an average mean squared error (MSE) of $1.62$~IMF\,\% and average mean absolute error (MAE) of $1.09$~IMF\,\%. The regression model fit of trial~2 is shown in Fig.~\ref{fig:SVR}. The high $R^2$ value of $0.94$ also indicates good prediction and generalisation performance of the developed SVR model.

\begin{table}[t]
  \centering
    \begin{tabular}{c|ccc|ccc}
	  & 
	  \multicolumn{3}{c|}{Model (n=55)} &
	  \multicolumn{3}{c}{Prediction (n=15)} \\
	  \cline{2-7}
	  Trial & ${R}^2$ & MSE & MAE & ${R}^2$ & MSE & MAE\\ 
	  \hline
	  1&0.93&1.10\,\%&0.79\,\%&0.93&1.46\,\%&1.02\,\%\\ 
	  \hline
	  2&0.94&1.23\,\%&0.91\,\%&0.93&2.27\,\%&1.22\,\%\\ 
	  \hline
	  3&0.91&1.17\,\%&0.83\,\%&0.95&1.22\,\%&0.88\,\%\\ 
	  \hline
	  4&0.92&1.42\,\%&0.94\,\%&0.97&1.32\,\%&0.87\,\%\\  
      \hline
	  5&0.93&0.98\,\%&0.70\,\%&0.90&1.84\,\%&1.48\,\%\\ 
	  \hline
	\end{tabular}
  \caption{Accuracy of the SVR prediction model for different trials. MSE and MAE stands for, respectively, mean squared and mean absolute errors, expressed in IMF\,\%.}
  \label{tab:SVR}
\end{table}

\section{Discussion}

\noindent IMF absorbs 1300~nm wavelength light stronger than muscle or water. When imaging meat samples with OCT, this difference can be detected in the OCT signal, enabling tissue discrimination and measurement of IMF\,\%. We have shown that an SVR prediction model of IMF\,\% based on the OCT data has an ${R}^2$ value of $0.94$ and an average mean absolute error of $1.09$~IMF\,\%, making it a promising tool to predict meat quality. The prediction accuracy of our OCT-based SVR model is better or as good as the models developed using various other techniques such as near infrared~(NIR) reflectance spectroscopy~(\cite{prieto2009application, rodbotten2000prediction, prieto2014use, su2014development}), hyper-spectal imaging~(\cite{elmasry2013chemical,kobayashi2010near}), Raman spectroscopy~(\cite{fowler2015prediction,andersen2018predicting}), X-ray computed tomography~(\cite{prieto2010predicting}, or bioelectrical impedance spectroscopy~(\cite{altmann2006prediction, marchello1999bioelectrical,damez2013quantifying}), and which have a prediction ${R}^2$ value in the range $0.48$--$0.98$. OCT however has the potential to enable in-line non-destructive testing of individual samples at a comparatively low cost, which none of these other techniques can provide. 

NIR spectroscopy is the most extensively used and studied technique for meat quality assessment due to its low cost and high fat sensitivity. Like other spectroscopic techniques, including Raman and bioelectrical impedance spectroscopy, it works best with minced meat. However, sampling uncertainties typically lead to large prediction errors when considering intact heterogeneous meat samples~(\cite{cheng2015marbling}). Spectroscopic techniques also all have the drawback of providing little or no  spatially-resolved information about the samples. Hyper-spectral imaging, which combines advantages of spectroscopic and imaging techniques, partially addresses that issue. However, long data processing times hinder real time applications. Additionally, the use of an external light source makes calibration across processing plants difficult in the long term. All these techniques are also essentially limited to surface examination, and cannot probe deep into meat samples. In comparison, OCT, which can probe beneath the surface of the meat samples with high resolution, leads to a high prediction accuracy even with intact samples. In addition, data acquisition and data analysis can be done in real time.

Freezing and thawing of the samples used in our OCT study results in water loss. This loss alters the myofibrillar protein structure and optical properties of meat muscle~(\cite{leygonie2012impact, xia2009physicochemical}), but does not change the fat content. Therefore this has no influence on the IMF\,\% prediction accuracy of our model.

Our study had access only to a limited number of samples, with a reduced spread of IMF\,\%. We therefore anticipate that a test of our method with more samples along the entire IMF\,\% range, especially towards high IMF content, is required before using it in an industry environment. To provide a wider variability of testing conditions, it would also be desirable to use samples from a broader range of breeds, muscle types, and diet.  

We must finally note that imaging a single side of large samples would be more suitable for an industry environment than imaging all sides of small samples. Large meat samples could be imaged in one single scan using a long-range and wide field of view OCT system (\cite{song2016long,shirazi2016fast}). Our current system takes $\approx$~55~seconds to image and analyse one sample with a 225~$\mathrm{cm}^2$ area, but this could be substantially reduced to 10~seconds or less, in particular by using recent multi-megahertz sweep rate sources (\cite{xu2015high, lee2015wide}). Our OCT technique will also need to be tested, and the prediction model rebuilt, in an abattoir atmosphere which presents additional industrial noise and challenges. This is one of the future development direction of this study.   

\section{Conclusion}

\noindent The work presented here brings insights on the ability of OCT and machine learning techniques to predict the quality of meat. To the best of our knowledge, this is the first time that OCT has been applied to this problem. We have shown that OCT has the potential for completely automated, contact-less, non-destructive, real time detection and classification of the quality of intact meat samples. 

\section{Acknowledgements}

\noindent The project was funded by the New Zealand Ministry of Business, Innovation and Employment (MBIE) under grant ``Capturing the value of red meat'' (C10X1602) as well as by the Marsden Fund of the Royal Society of New Zealand. We also acknowledge Drs~Talia Maree Hicks and Cameron Craigie from Agresearch for providing samples for this study.   

\section*{References}
\bibliography{mybibfile}

\end{document}